\documentclass[aps,pra,twocolumn,superscriptaddress,nofootinbib]{revtex4-2}

\usepackage[dvipsnames]{xcolor}
\usepackage{hyperref}
\hypersetup{
  breaklinks=true,
  colorlinks=true,
  allcolors=BlueViolet,
}

\usepackage{graphicx} 

\usepackage{physics,braket,bm}
\renewcommand{\set}[1]{\left\{#1\right\}}

\newcommand \sfI{{\sf I}}

\newcommand \sfR{{\sf R}}

\newcommand \sfU{{\sf U}}

\newcommand \be{\begin{equation}}
\newcommand \ee{\end{equation}}

\newcommand \bea{\begin{eqnarray}}
\newcommand \eea{\end{eqnarray}}

\newcommand{\ord}[1]{\mathrm{ord}(#1)}

\usepackage{amssymb}

\usepackage[inline]{enumitem}
\setlist[enumerate]{leftmargin=*}

\newcommand{\InfleqtionC}{Infleqtion, Inc., Chicago, IL, 60615, USA}
\newcommand{\InfleqtionM}{Infleqtion, Inc., Madison, WI, 53703, USA}
\newcommand{\UWM}{Department of Physics, University of Wisconsin-Madison, 1150 University Avenue, Madison, WI, 53706 USA}

\definecolor{mscolor}{rgb}{0,0.5,0.5}
\definecolor{cpcolor}{rgb}{0.4,0,0.8}
\definecolor{tgcolor}{rgb}{0.3,0.6,0.3}

\usepackage{comment}

\newcommand{\rsub}[1]{\textcolor{black}{#1}}

\begin{document}

\title{Architecture for fast implementation of qLDPC codes with optimized Rydberg gates}
\author{C. Poole}
\affiliation{\UWM}
\author{T. M. Graham}
\affiliation{\UWM}
\author{M. A.~Perlin}
\affiliation{\InfleqtionC}
\author{M. Otten}
\affiliation{\UWM}
\author{M. Saffman}
\affiliation{\UWM}
\affiliation{\InfleqtionM}
\date{\today}

\begin{abstract}
We propose an implementation of bivariate bicycle codes (Nature {\bf 627}, 778 (2024)) based on long-range Rydberg gates between stationary neutral atom qubits. An optimized layout of data and ancilla qubits reduces the maximum Euclidean communication distance needed for non-local parity check operators. An optimized Rydberg gate pulse design enables $\sf CZ$ entangling operations with fidelity ${\mathcal F}>0.999$ at a distance greater than $12~\mu\rm m$. The combination of optimized layout and gate design leads to a quantum error correction cycle time of $\sim 1.2\rsub{8}~\rm  ms$ for a
$[[144,12,12]]$ code,  \rsub{nearly a factor of two }
improvement over previous designs. 
  
\end{abstract}

\maketitle

\section{Introduction}

Recently, a new family of quantum low density parity check (qLDPC) codes known as quasi-cyclic ``bivariate bicycle'' codes was proposed and analyzed in Ref.~\cite{Bravyi2024}.
These  bicycle codes were found to have encoding rates greatly exceeding that of the surface code \cite{Fowler2012}, pseudo-thresholds competitive with the surface code, and a Tanner graph with ``thickness'' 2.
As pointed out in Ref.~\cite{Bravyi2024}, the thickness-2 Tanner graph motivates a qubit layout on a dual-sided superconducting chip. 
The bicycle codes have weight-6 check operators whose measurement requires long-distance interactions.
The need to incorporate long-distance connections is intrinsic to any \rsub{planar} code that improves on the density of stored quantum information compared to the surface code, which saturates the BPT bound found in Ref.~\cite{Bravyi2010}.

It has been proposed \cite{Viszlai2023} to implement the non-local connectivity required for bivariate bicycle codes with neutral atom qubits using atom-transport techniques \cite{Bluvstein2024}.
While transport provides long-range connectivity, it is also intrinsically limited in speed \cite{Lam2021}, and quantum error correction (QEC) cycle times for measuring a full set of $\sf X$ and $\sf Z$ parity checks were estimated to be $\sim \rsub{2}$ ms. An implementation of hypergraph qLDPC codes based on atom transport also leads to QEC cycle times of tens of milliseconds \cite{QXu2024}, \rsub{albeit for a larger qubit count but also higher logical error rate than we consider in the following}.  
This slow cycle time calls into question the utility of these implementations for applications that require deep circuits or large numbers of samples \cite{Beverland2022}.

\begin{figure}[!t]
\centering
 \includegraphics[width=\columnwidth]{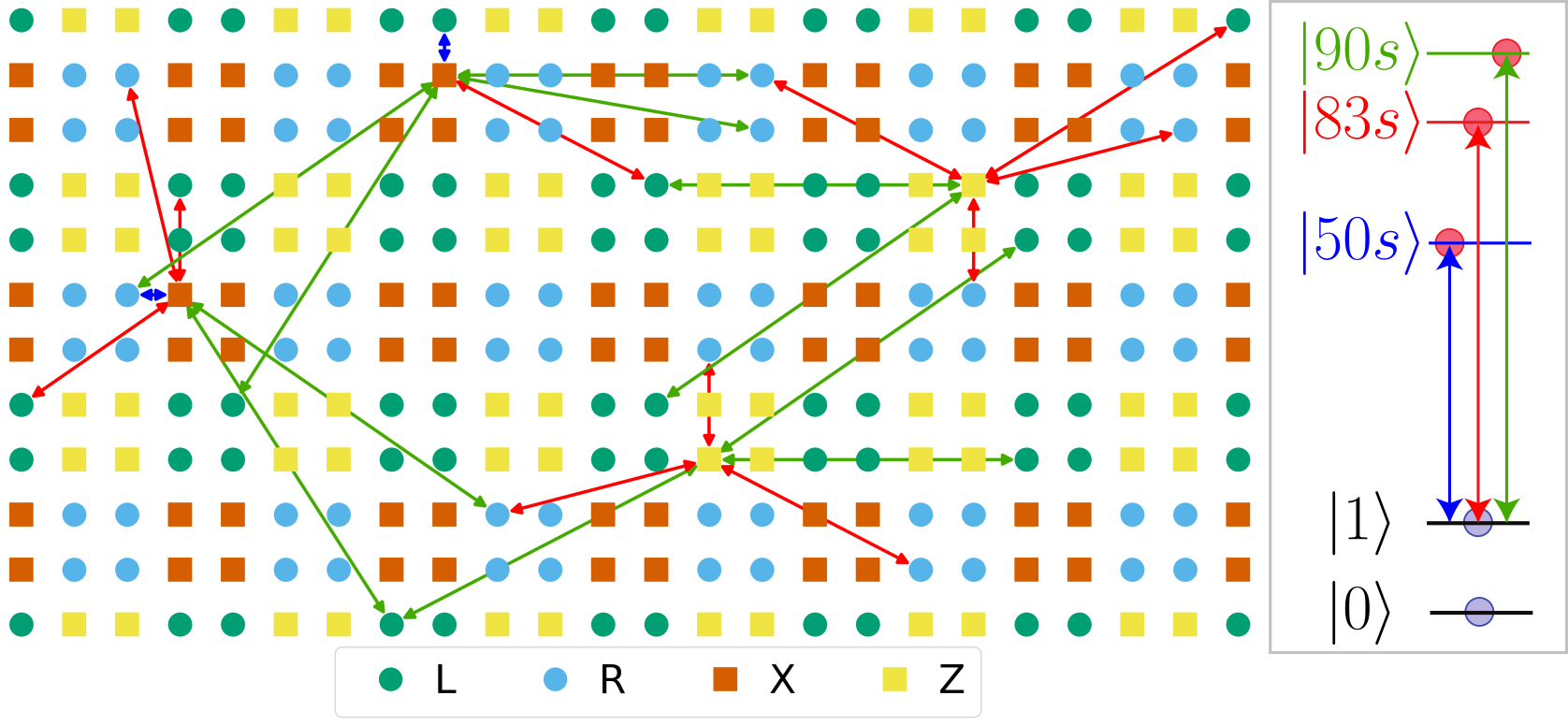}
 \vspace{-.6cm}
  \caption{Optimized layout of the [[144,12,12]] code with a maximum communication distance of 7.21 lattice spacings.  The check operations are shown for a few ancilla qubits with color coded connections corresponding to the Cs Rydberg levels shown on the right. The qubit color coding is $L$ data = green circle, $R$ data = blue circle,  $\sf X$ check = red square, $\sf Z$ check = yellow square.}
\label{fig.layout}
\end{figure}

We propose here to use a neutral atom architecture based on fast optical beam scanning \cite{Graham2022} that incorporates two innovations to reduce the QEC cycle time by \rsub{nearly a factor of two } compared to transport-based approaches. \rsub{We define cycle time as the time required to implement a full set of $\sf X$ and $\sf Z$ stabilizer gate operations, which is the definition used in related work\cite{Viszlai2023}.  The actual QEC cycle rate includes additional time to reset check qubits and to measure them. Estimates of how these additional operational times impact the cycle time are included in Sec. \ref{sec.time}.}  
Our first contribution, illustrated in Fig.~\ref{fig.layout}, is the introduction of a qubit layout that can reduce the largest Euclidean communication distance that is required to perform parity check operations by almost a factor of 3 relative to known layouts \cite{Bravyi2024, Viszlai2023}. The second   contribution is an optimized Rydberg gate design that minimizes the required two-atom interaction strength while maintaining fast speed and high fidelity. Together, these approaches make it feasible to implement high performance bicycle codes up to  distance $d=18$ without relying on atom transport. The estimated QEC cycle times, assuming fast qubit measurements, are $\sim1.2\rsub{8}~\rm ms$, which is \rsub{nearly a factor of two }
improvement over existing proposals. We address the issue of achieving fast qubit measurement time in Sec. \ref{sec.time}. 

\section{Optimizing the qubit layout}

\begin{figure}[!t]
\centering
 \includegraphics[width=.95\columnwidth]{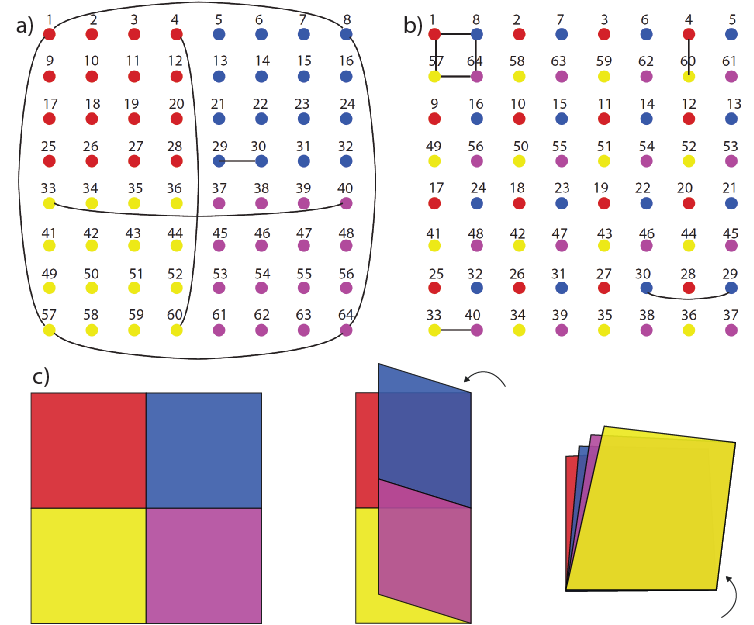}
  \caption{
  a) A grid of nearest-neighbor connected qubits on a torus has long range connections when mapped onto a planar grid with open boundary connections.
  b) A dilation and remapping procedure converts long-range connections into nearest-neighbors, and neighboring connections into nearest- or next-nearest neighbors \cite{EMa1993}.
  c) This remapping is achieved by ``folding'' the plane in a) once vertically and again horizontally and expanding each quadruplet of overlapping qubits into a square as in panel b).
  Only a few connections are shown for clarity.
  }
\label{fig.torus}
\end{figure}

Table \ref{tab.qLDPC} provides, for several bivariate bicycle codes with code distances $d=6$--$18$, the maximum  communication distance $D_{\rm max}$ required to implement these codes when qubits are laid out on a rectangular grid as described in Ref.~\cite{Bravyi2024} and \cite{Viszlai2023}, as well as the required $D_{\rm max}$ when following a qubit assignment procedure that we describe below. 
For several of the codes, including the $[[144,12,12]]$ code that we will examine in more detail, the maximum distance arises from the check operation which would be nearest neighbor on a torus, but stretches from one side of the qubit grid to the other for a planar layout with open boundary conditions, as in Fig. \ref{fig.torus}a.

\begin{table}
\caption{
Bivariate bicycle codes with $n$ data qubits, $2n$ total qubits (including ancillae), $k$ logical qubits, and code distance $d$.
The figure of merit $r=k d^2/n$ quantifies an improvement over the (rotated) surface code for which $r=1$ with any choice of $k$ or $d$.
For each code, we report the maximum   communication distance $D_{\rm max}$ in units of the lattice spacing using the qubit layouts in Refs.~\cite{Bravyi2024, Viszlai2023}, and with optimized layouts found in this work (see main text for explanation).
}
\label{tab.qLDPC}
\centering
\begin{tabular}{|c| c| c|c|c|c|}
\hline
code & f.o.m. & IBM \cite{Bravyi2024} & UC \cite{Viszlai2023} & this work  \\
$[[n,k,d]]$ & $r$ & $D_{\rm max}$ & $D_{\rm max}$ & $D_{\rm max}$  \\
\hline
$[[72,12,6]]$ & 6. & 11. & 11. & 5.  \\
$[[90,8,10]]$ & 8.89 &29. & 25. & 10. \\
$[[108,8,10]]$ & 7.41 &17. &15. & 7. \\
$[[144,12,12]]$ & 12. & 23. &21. & 7.21 \\
$[[288,12,18]]$ & 13.5 &27.7 & 21.& 7.21\\
\hline
\end{tabular}
\end{table}

Rydberg gates for neutral atom qubits require an atom spacing that is at least the diameter of the Rydberg wavefunction in order to avoid collisions between Rydberg electrons or between a Rydberg electron and a neighboring atomic core.  This limit is about $1~\mu\rm m$, which is the wavefunction \rsub{radius} for a Rydberg atom in a low angular momentum state with principal quantum number $n=100$. There are also practical limits to how tightly the laser beams used for Rydberg excitation can be focused without crosstalk arising from tails of the optical beam profiles. A feasible, albeit demanding, practical limit is an atom separation of at least $s_{\rm min}\simeq 1.5~\mu\rm m$.  A high fidelity Rydberg $\sf CZ$ gate requires at least a few MHz of interaction strength \cite{Jandura2022}, and this  limits the physical interaction distance to $R_{\rm max}=10-15~\mu\rm m$, or about $10 s_{\rm min}.$  We see that the layouts given in Refs.~\cite{Bravyi2024, Viszlai2023} only approach this limit for the $[[72,12,6]]$ code, but have substantially longer communication distances for the higher performance codes. 

In order to implement check operations without atom transport, the communication distance, or the requirement on Rydberg interaction strength, needs to be reduced. We proceed to show how both the distance and the requirement on Rydberg interaction can be improved. To reduce the communication distance, we change the layout of the qubits in a planar rectangular grid, as in Fig.~\ref{fig.layout}. We proceed by placing
\rsub{all}
qubits in the plane \rsub{by} following an enumeration scheme
\rsub{that generalizes}
the method used in \cite{Bravyi2024}, but which is not restricted by the requirement that there are four nearest neighbor check operators for each qubit under periodic boundary conditions \rsub{(see Appendix \ref{sec.App.layout})}.
The entire grid is then remapped according to the  ``\rsub{dilation and} folding" procedure illustrated in Fig. \ref{fig.torus}\rsub{\cite{EMa1993}}. At this stage, the positions of $L$ and $R$ qubits are considered fixed, and
\rsub{we re-map $\sf X$ and $\sf Z$ check qubits in such a way as to minimize}
the maximum \rsub{communication} distance $D_{\rm max}$.

\rsub{
Our general strategy for minimizing the maximum communication distance $D_{\rm max}$ is to (a) identify, for a given $D_{\rm max}$, whether there exists \emph{any} placement of $\sf X$ and $\sf Z$ qubits such that every qubit is at most $D_{\rm max}$ away from the qubits that it must interact with, and then (b) incrementally decrease $D_{\rm max}$ until no such placement exists.
To this end, for any given $D_{\rm max}$ we build a bipartite graph in which the first partition of $n$ nodes represents the $\sf X$ and $\sf Z$ qubits, and the second partition of $n$ nodes represents unoccupied lattice sites in which to place these qubits.
For every qubit $q$ and lattice site $\ell$, we add the edge $\set{q,\ell}$ to the graph if $q$ is at most a distance $D_{\rm max}$ away from the $L$ and $R$ qubits that it must address when $q$ is placed at $\ell$.
A perfect matching of this graph---if it exists---is then an assignment of qubits to lattice sites in such a way that the maximum communication distance is at most $D_{\rm max}$.
The existence of a perfect matching can be determined efficiently (in polynomial time) by using the Hopcroft-Karp algorithm \cite{Hopcroft1973} to find the cardinality of the maximum matching of the graph.
To find the minimum satisfiable $D_{\rm max}$, we enumerate all possible values of $D_{\rm max}(\delta x, \delta y) = \sqrt{ \delta x^2 + \delta y^2}$ and test progressively more restrictive values until a perfect matching cannot be found.
}
We search over a broad range of fixed data qubit layouts to find the conditions that give the smallest communication distances given in Table \ref{tab.qLDPC}. A description of how the data qubit layouts in the search were generated is provided in Appendix \ref{sec.App.layout}.
We see that this procedure reduces the maximum communication distance for all codes in Table \ref{tab.qLDPC} and for the $[[144,12,12]]$ and $[[288,12,18]]$ codes the reduction is almost a factor of three. Four of the five codes after remapping have $D_{\rm max}\le 7.21$ which is achievable with Rydberg $\sf CZ$ gates that have been optimized for long distance operation.

\begin{figure}[!t]
\vspace{-.0cm}
\centering
 \includegraphics[width=8.5cm]{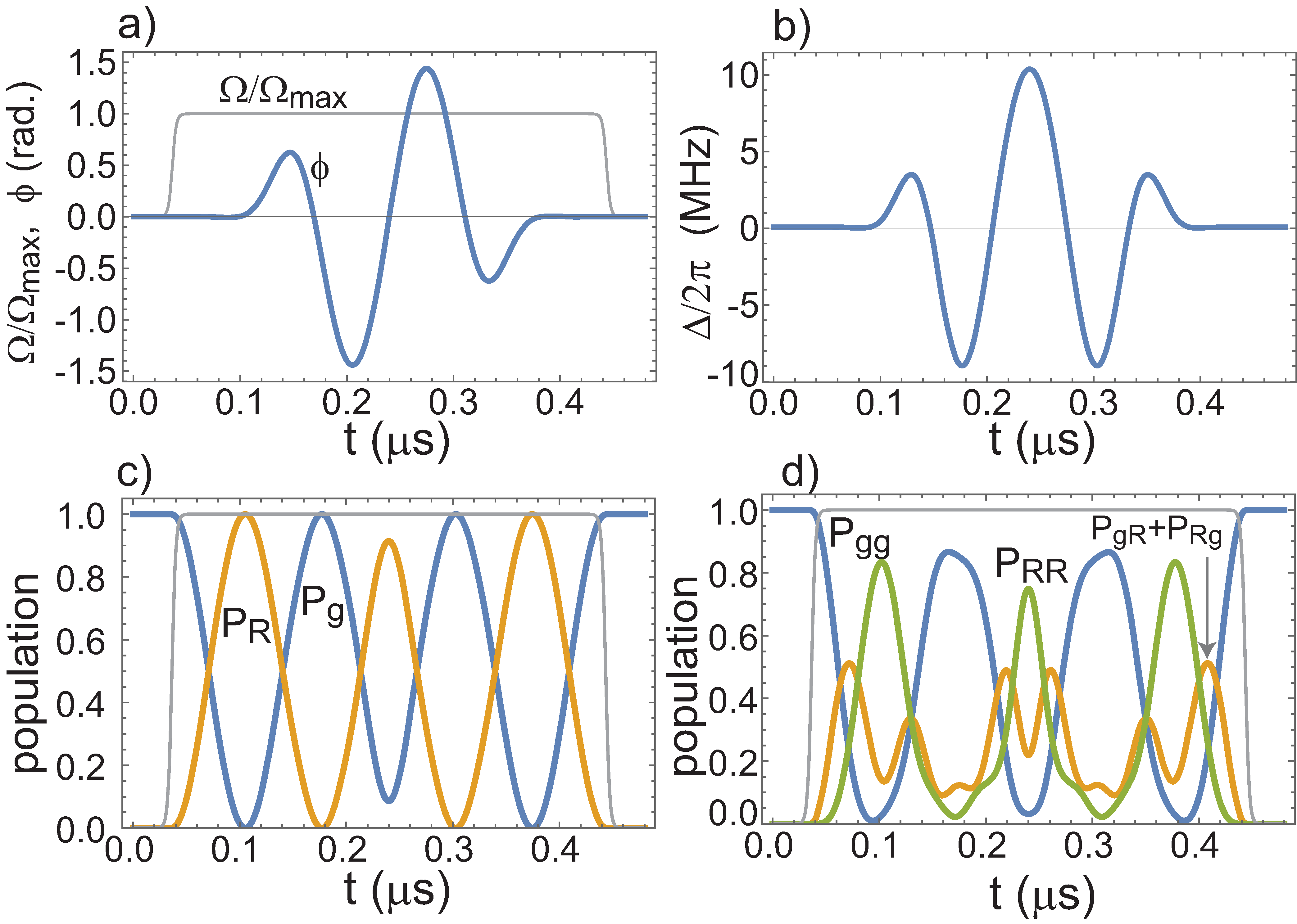}
\vspace{-.3cm}
  \caption{Rydberg gate simulation at low interaction strength $V/(2\pi)=3.8~\rm MHz$ and Rydberg lifetime $\tau_{\rm R}=252~\mu\rm s$, corresponding to Cs $90s_{1/2}$ giving error $\epsilon=7.\times 10^{-4} $. Panels show (a) Rabi drive \rsub{$(\Omega_{\rm max}/2\pi=7.3~\rm MHz)$} and phase profile, (b) corresponding detuning, (c) ground $P_{\rm g}$ and Rydberg $P_{\rm R}$ populations for one atom coupled to the Rydberg state, and (d) ground $P_{\rm gg}$, singly excited $P_{\rm gR}+P_{\rm Rg}$ and double Rydberg $P_{\rm RR}$ populations for two atoms coupled to the Rydberg state. Gate parameters are given in row 17 of Table \ref{tab.distances2} in \rsub{Appendix \ref{sec.App.gate}}. }
\label{fig.optimal}
\end{figure}

\section{Optimized Rydberg gates}

There is a large literature on analysis and design of  protocols for entangling Rydberg 
gates \cite{Jaksch2000, Saffman2010, XZhang2012, Levine2019, Robicheaux2021, Jandura2022}. 
It is useful to distinguish between gate errors that arise due to various technical limitations (e.g., laser noise, optical beam pointing,
atom motion and position variations at
finite temperature,
electric and magnetic field noise) and fundamental limits set by atomic structure and quantum physics. Ignoring technical limitations, the entanglement fidelity $\mathcal F$ of a Rydberg-mediated $\sf CZ$ gate is fundamentally limited by
\be
{\mathcal F}_{\rm max}\le 1 - \frac{2}{V \tau_{\rm R}} ~~~{\rm or}~~~ \epsilon_{\rm min}=1-{\mathcal F}_{\rm max}=\frac{2}{V \tau_{\rm R}} ,
\label{eq.fidelitylimit}
\ee
where $\epsilon_{\rm min}$ is the minimum gate error, $V$ is the interaction strength between two Rydberg excited atoms, and $\tau_{\rm R}$ is the Rydberg radiative lifetime. This limit follows from the analysis in Ref.~\cite{Wesenberg2007}, and governs the fidelity that can be achieved when two particles interact via states with finite lifetimes. This limit is independent of the specific laser pulse protocol that is used to implement an entangling gate, and thus provides a benchmark to compare entanglement protocols against.

Recent progress in high fidelity Rydberg gates leading to fidelities of ${\mathcal F}> 0.99$ \cite{Evered2023,Peper2024,Finkelstein2024,Radnaev2024,Muniz2024} has relied on a time optimal protocol that uses a single shaped pulse addressing both atoms \cite{Jandura2022,Pagano2022,Mohan2023}. With correct design of the pulse shape, the gate error is dominated by the amount of Rydberg scattering, which can be written as $\epsilon_{\rm R}=T_{\rm R}/\tau_{\rm R}$, where $T_{\rm R}$ is the  integrated Rydberg population during the gate and $\tau_{\rm R}$ is the Rydberg state lifetime.  In
\rsub{Fig.~7 of Ref.~\cite{Jandura2022} gate designs were presented for finite blockade strength  which reach within a factor of about \rsub{15} of $\epsilon_{\rm min}$ under conditions of strong blockade.}  
Other gate designs \cite{Petrosyan2017} have reached similar results of $\epsilon\simeq 19\epsilon_{\rm min}$, which is also far from the fundamental limit of Eq.~\eqref{eq.fidelitylimit}.

We have improved upon this limit reaching $\epsilon= \frac{\rsub{3.4}}{V \tau_{\rm R}}$, a factor of \rsub{1.7} from the fundamental limit,  using an analytical pulse shape as shown in
Fig. \ref{fig.optimal}. The Rabi pulse has constant amplitude with smooth turn on and turn off edges. The phase profile of the pulse is 
\be
\phi(t)=\Delta_0 t + a\sin[2\pi f(t-t_0)]e^{-[(t-t_0)/\tau]^4}
\ee
with $\Delta_0$ the detuning from ground-Rydberg resonance, $a$ the amplitude of the phase modulation, $f$ the modulation frequency, $t_0$ the midpoint of the pulse, and $\tau$ providing an envelope width. This phase profile is similar to that in Ref.~\cite{Evered2023}, with the addition of the super-Gaussian envelope and enforcing of anti-symmetry of the phase about the pulse midpoint. The numerical simulation in Fig. \ref{fig.optimal} accounting for Rydberg decay, but not including any of the technical noise sources enumerated above, verifies that ${\mathcal F}>0.999$ can be achieved with a relatively small interaction strength of 3.8 MHz and a Rydberg lifetime of $\tau_{\rm R}=252~(\mu\rm s)$, which corresponds to the Cs $90s_{1/2}$ state 
in a room temperature environment. 
Details of the simulation methodology and calculation of the fidelity are given in \rsub{Appendices \ref{sec.App.fidelity} and \ref{sec.App.gate}. The parameters for each gate used in the [[144,12,12]] code can be found in table \ref{tab.distances2}.}

 \begin{table}[!t]
\caption{
Distribution of  communication distances $D$ in units of lattice spacing for the 864 check operations in the   $[[144,12,12]]$ code. The interaction strength is given at  $R=1.7 D~ (\mu\rm m)$. The lifetimes of the Rydberg levels are $\tau_{\rm R}=60.4, 209, 252~(\mu\rm s)$ for $n=50,83,90.$ Details of the Rydberg gate parameters are given in \rsub{Appendix \ref{sec.App.fidelity}}. 
\rsub{The listed fidelities $\mathcal F'$ include errors from Rydberg-Rydberg forces and Doppler effects from Table \ref{tab.dipole_dipole}.}
} \label{tab.distances}
\centering
\begin{tabular}{|c| c| c|c|c|c|c|}
\hline
gate&distance & occurrences & level & interaction & fidelity & $t_{\rm gate}$ \\
&$D$& & & $\frac{V(R)}{2\pi}$ (MHz)  & $\mathcal F'$ & (ns)\\
\hline
1&1 & 80 & $50s_{1/2}$&415. & 0.9996 & 130\\
2&1.41 & 4& $50s_{1/2}$&58.5 & 0.9991 & 180\\
\hline
3&2 & 88& $83s_{1/2}$&1160. & 0.9999 & 150 \\
4&2.24 & 16& $83s_{1/2}$&780. & 0.9999 & 150\\
5&3.16 & 12& $83s_{1/2}$&170. & 0.9998 & 180\\
6&3.61 & 12& $83s_{1/2}$&85. & 0.9998 & 180 \\
7&4.12 & 56& $83s_{1/2}$&40. & 0.9997 & 180\\
8&4.47 & 112& $83s_{1/2}$&25. & 0.9994 & 200\\
9&5 & 24& $83s_{1/2}$&13. & 0.9994 & 270\\
10&5.10 & 4& $83s_{1/2}$&11.5 & 0.9993 & 270\\
11&5.83 & 8& $83s_{1/2}$&5.2 & 0.9991 & 270\\
\hline
12&6 & 112& $90s_{1/2}$&11.2 & 0.9994 & 350\\
13&6.08 & 48& $90s_{1/2}$&10.1 & 0.9992 & 355\\
14&6.40 & 16& $90s_{1/2}$&7.7 & 0.9991 & 430\\
15&6.71 & 52& $90s_{1/2}$&5.8 & 0.9992 & 440\\
16&7.07 & 4& $90s_{1/2}$&4.25& 0.9991 & 480\\
17&7.21 & 216& $90s_{1/2}$&3.8& 0.9990 & 480\\
\hline
\end{tabular}
\end{table}

We note that the Rabi frequency used in Fig.~\ref{fig.optimal} is similar to that which provided an experimental fidelity of 0.995 in Ref.~\cite{Evered2023}, but the interaction strength is more than 100 times smaller. This reduction in interaction strength 
is significant because it enables a large increase in two-atom separation.  With  the Cs $90s_{1/2}$ state an interaction strength of $V/(2\pi)=3.8~\rm (MHz)$ is reached at $R_{\rm max}=12.3~(\mu\rm m)$ which implies a bicycle code with communication distance of $D=7.21$ can be implemented with an array of atoms spaced by $s=1.7~(\mu\rm m)$.

\rsub{
The fundamental limit given in Eq. (\ref{eq.fidelitylimit}) only includes infidelity due to gate rotation errors and Rydberg decay; however, other error sources are relevant for long-range Rydberg gates. In such gates, the Rydberg blockade mechanism is not strong
which  makes double excitation of both atoms to the Rydberg state a more relevant consideration. When both atoms are excited to the Rydberg state, dipole-dipole repulsion causes dephasing errors, leading to infidelity \cite{Robicheaux2021}. To minimize such errors during long-range Rydberg gates, one must limit the Rabi frequency thereby increasing the gate time. This longer gate time increases the time each atom spends in the Rydberg state, increasing the relevance of infidelity due to the Doppler effect \cite{Robicheaux2021}. In Table \ref{tab.dipole_dipole}, in Appendix \ref{sec.App.gate}, we calculate the gate fidelity including double Rydberg excitations and the Doppler effect in addition to gate rotation errors and Rydberg decay.}

\section{QEC cycle time}
\label{sec.time}

We turn now to estimation of the time needed for each  QEC cycle. This estimate depends crucially on the concurrency of the syndrome check operations. 
While neutral atom architectures support long-range qubit interactions, which are critical for qLDPC bicycle codes,
the long-range nature of the Rydberg interaction limits the concurrency, forcing us to execute only one two-qubit gate at a time within a geometrical exclusion zone \cite{Baker2021}. At long range, the interaction strength decays as $V(R)\sim 1/R^6$ \cite{Walker2008}. 
\rsub{We performed simulations of the sensitivity of the $\sf CZ$ gates to blockade crosstalk as a function of $x_i = \Delta_{{\rm B}i} t_{\rm gate} / (2 \pi)$, where $\Delta_{{\rm B}i}$ is the sum of the interaction strengths between atoms not in the same $\sf CZ$ pair as qubit $i$, with qubit $i$. For all gate distances, if both qubits satisfy $x_i < 0.01$, then the infidelity due to blockade leakage is upper bounded by $10^{-4}.$}
The $[[144,12,12]]$ code uses 288 \rsub{data and check} qubits laid out on a $12\times 24$ grid  with a corner to corner distance of 26.8 lattice units.  The remapping procedure leads to 17 different check operator distances, each of which occurs the number of times shown in Table \ref{tab.distances}.  In order to increase the concurrency and reduce the cycle time, the shorter range gate operations are implemented with lower Rydberg levels that have a smaller exclusion distance.

We estimate the time needed for a full cycle of 864 check operations for the $[[144,12,12]]$ code using the distances and gate times in Table \ref{tab.distances}. \rsub{The check circuit consists of $\sf CZ$ operations, parallelized single qubit phase gates to correct local phases from the Rydberg pulse implementation of the $\sf CZ$ gates, and a combination of global microwave and parallelized Raman gates to perform basis transformations. In total, these local gates incur approximately ~$3.5~\mu\rm  s$ of additional gate time and four units of beam switching time. For the $\sf CZ$ gates, we assume that only gates with the same qubit distance can be parallelized to avoid incurring additional system complexity.} 
Given the above exclusion zone constraint, the \rsub{448 }
checks in rows \rsub{12}
-17 must be performed one at a time for our layout. For rows 1-\rsub{11}
, the \rsub{416 }
operations can be at least partially parallelized, allowing for a reduction \rsub{of 171 }
cycles giving a total of \rsub{693 }
\rsub{to implement all $\sf CZ$ gates.} Details for achieving this reduction are given in \rsub{Appendix \ref{sec.App.concurrency}}. Accounting for this reduction the effective average \rsub{{\sf CZ} illumination }
time  is $\langle t_{\rm gate}\rangle \simeq\rsub{0.27}
~\mu\rm s$. \rsub{In total, this leads to $237.5~\mu\rm s$ of fixed illumination time plus 697 units of switching overhead.}
\rsub{W}e allow for  a switching time between optical beam configurations of $1.5~\mu\rm s$ based on proposed architectures for fast photonic beam patterning \cite{Graham2023,Menssen2023,BZhang2024}, which leads to a cycle time for a full set of $\sf X$ and $\sf Z$ check operations of \rsub{1.28 \rm ms}
. This value is strongly dependent on assumptions about the optical control system and reconfiguration speed and could be reduced by multiplexing multiple spatial light modulators, at the cost of additional system complexity. We anticipate that the requisite optical technology will continue to advance and lead to faster execution times, whereas atom transport cannot be accelerated to arbitrary speeds given quantum speed limits for trapped atoms \cite{Lam2021}.

\begin{figure}[!t]
\centering
 \includegraphics[width=.45\textwidth]{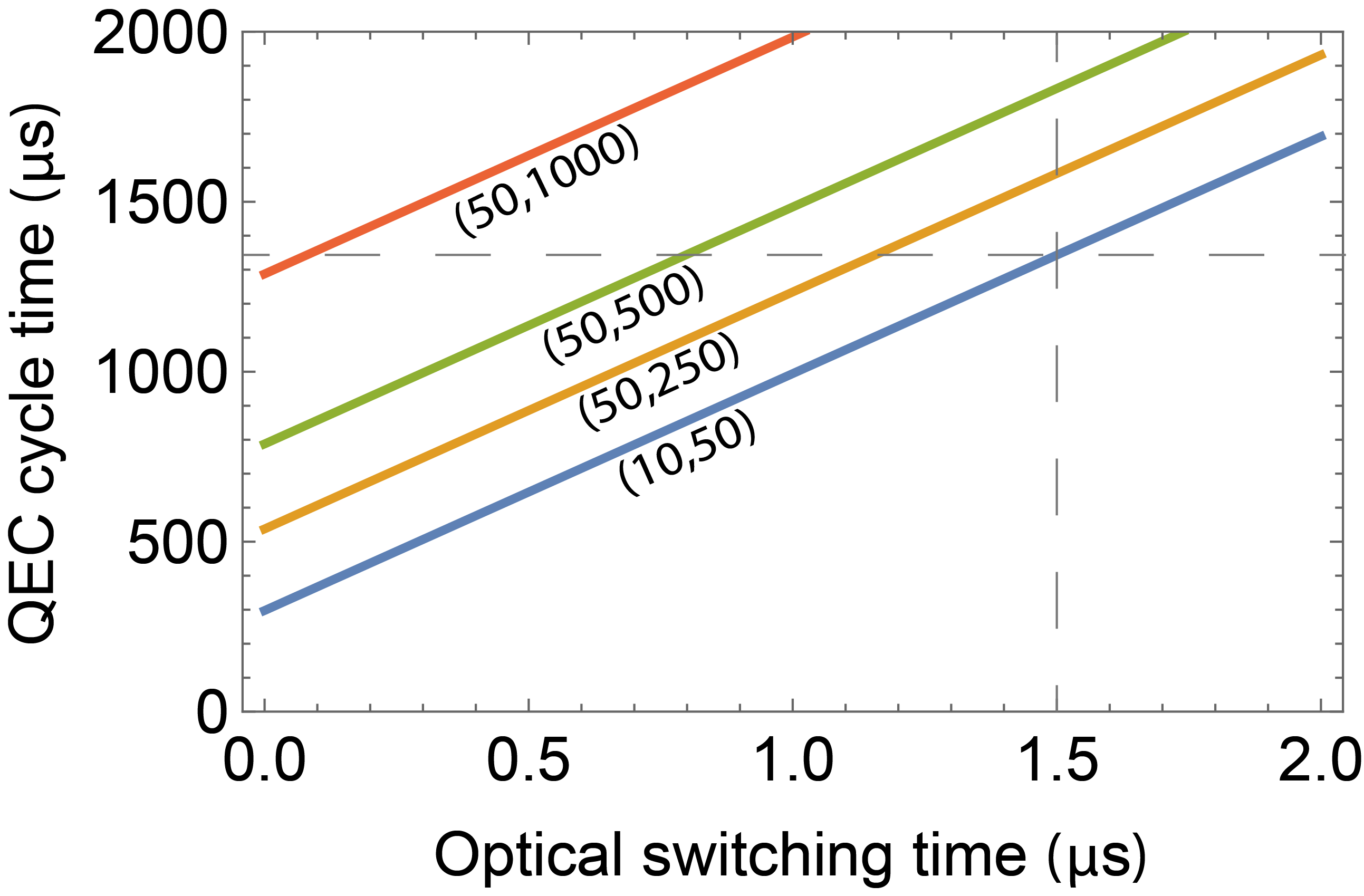}
 \vspace{-.0cm}
  \caption{\rsub{Duration of a single round of QEC for the optimized layout of the $[[144,12,12]]$ code as a function of the overhead time due to optical beam switching. The different curves include the time for  optical pumping for ancilla reset ($t_{\rm op}$) and ancilla measurement ($t_{\rm meas}$). The curves are labeled with $(t_{\rm op},t_{\rm meas})$ in $\mu\rm s.$ The dashed lines show the cycle time with the assumption of $1.5~\mu\rm s$ switching time assumed in the main text. 
 }}
\label{fig.QECtime}
\end{figure}

\rsub{Figure \ref{fig.QECtime} shows the dependence of the cycle time for the [[144,12,12]] code on the time to reconfigure the optical gate control beams. For codes other than the [[144,12,12]] code, we can place an upper bound on the time by assuming no {\sf CZ} parallelization is possible and all {\sf CZ} gates have worst case illumination time $t_{\rm max}$, resulting in a time at most $t\le6N(t_{\rm max}+t_{\rm switch})+4t_{\rm switch}+3.5~\mu\rm s$. Here, $t_{\rm switch}$ is the switching overhead time. For the [[144,12,12]] code, $t_{\rm max}=0.48 ~\mu\rm s$ and we have assumed $t_{\rm switch}=1.5~\mu\rm s$, leading to an upper bound of $1.72~\rm  ms$. Our actual estimate accounting for the parallelization scheme and the accurate illumination time of all the Rydberg gates is 1.28 ms, a 25\% savings relative to this upper bound.}

In addition to a fast rate of syndrome extraction operations, a fast QEC cycle requires fast measurement and ancilla reset operations.  
State measurements of arrays of neutral atom qubits are most often performed by analysis of fluorescence images acquired by illuminating the atoms with light that resonantly couples one of the qubit states to a short lived excited state, while the other qubit state is sufficiently far detuned to be dark to the imaging light. If the excited state has lifetime $\tau$, photons can be scattered at a maximum rate of $r=1/(2\tau)$. The rate of detected photons is $r_{\rm det}=\eta r$ where $\eta$ accounts for the solid angle of the imaging optics, losses in optical components, and detector quantum efficiency. Typical values for optimized alkali atom experiments are $\tau\sim 30~\rm ns$ and $\eta=0.1$ which leads to a detection rate that could  reach  $r_{\rm det}>  1\times 10^6~\rm s^{-1}$. 

Neglecting detector noise, a state measurement can be based on detection of a single photon, such that sub-$\mu$s measurements are  in principle possible. In practice, the maximum scattering rate is not used since it leads to excessive atom heating and detector noise necessitates longer integration times, such that state measurements typically require several ms \cite{Graham2023b}, although measurements as fast as 0.25 ms have been reported in free space \cite{Chow2023}. Measurement times can be reduced further using more sophisticated image analysis \cite{Phuttitarn2024} \rsub{or incorporation of auxiliary measurement atoms\cite{Petrosyan2024}}. Recently, atom occupancy measurements with integration times as short as $2.4 ~\mu\rm s$ have been achieved \cite{LSu2024}. While demonstration of state measurements in large arrays at few $\mu\rm s$ timescales is an outstanding challenge, we anticipate this will be achieved, such that the time needed for syndrome check operations remains the limiting factor for fast QEC operation.  \rsub{Figure \ref{fig.QECtime} shows the cycle time for several values of the reset and measurement times.  }

\rsub{The error syndrome circuit assumes each of the ancilla qubits is initialized to the $\ket{0}$ state. To ensure that this is the case for each round of check operations and measurement, there are a few available options. Each of the ancilla qubits measured to be in the $\ket{1}$ state could be selectively rotated back to $\ket{0}$ via parallelized Raman gates. Alternatively, if the ancilla and data qubits are different species\cite{Beterov2015}, all ancilla could be optically pumped back into the $\ket{0}$ state without affecting the quantum state of the data qubits \rsub{\cite{Singh2023}}. It is also possible to discard the ancilla qubits and load fresh atoms prepared in the $\ket{0}$ state into the corresponding traps. The time cost associated with each of these options would depend strongly on the specifics of the architecture.}

\section{Summary}

In summary, we have presented an implementation of qLDPC bivariate bicycle codes based on long range Rydberg gates, which has the potential for reducing the QEC cycle time by \rsub{nearly a factor of two } 
compared to an architecture based on atom transport. The achievable cycle time is critically dependent on the availability of fast photonic beam switching, and improved optical devices will enable further time reductions. 

\rsub{The bicycle and other qLDPC codes provide quantum memory with high code rate. Complete solutions for performing logic in place in these codes are still under development\cite{GZhu2023,QXu2024b}. Alternatively, hybrid architectures have been envisioned based on teleportation between a qLDPC memory and the surface code for logical operations\cite{Viszlai2023,QXu2024}. In such hybrid approaches the cycle rate of the code blocks performing logical operations  need not be the same as the QEC cycle rate in the qLDPC code. The ability to implement codes with high distance, such as the $d=12$ case considered here, coupled with the long coherence times of neutral atom qubits, makes these codes promising for space and time efficient memory. As long as there are sufficient memory blocks to keep the logical units supplied with data,  the qLDPC cycle time will not limit the rate of logical computation.  }
\\

\acknowledgments

We thank Denny Dahl for sharing the dilation method of planar mapping and Fred Chong for pointing us to Ref.~\cite{EMa1993}.
This material is based upon work supported by the
U.S. Department of Energy Office of Science, National
Quantum Information Science Research Centers as part of the Q-NEXT center, as well as support from NSF Award PHY-2210437 and NSF award 2016136 for the QLCI center Hybrid Quantum Architectures and Networks.

While completing this manuscript we became aware of related work on implementation of qLDPC hypergraph codes using Rydberg gates \cite{Pecorari2024}.

\bibliography{qc_refs,optics,saffman_refs,rydberg,atomic}

\appendix

\vspace{1.cm}

The appendices contain additional information concerning the procedure for optimizing the qubit layout, $\sf CZ$ fidelity calculation, simulations of Rydberg gate fidelity,  gate parameters, and concurrency of gate operations.

\section{Optimizing the qubit layout}
\label{sec.App.layout}

 The bivariate bicycle qLDPC codes are constructed in the following manner \cite{Bravyi2024}. Let $S_\ell$ be the cyclic shift matrix of dimension $\ell \times \ell$. Row $i$ of $S_\ell$ has a single nonzero entry equal to one at column $i\,{+}\,1{\pmod \ell}$. For example, 

 $$S_3=\begin{bmatrix}
 0 & 1 & 0 \\
 0 & 0 & 1 \\
 1 & 0 & 0 
\end{bmatrix}.$$

Note also that $S^\ell_\ell$ has order $\ell$, $S^\ell_\ell = I_\ell$ the $\ell\times \ell$ identity matrix. We can define polynomials  
\begin{equation}
    x^py^q=S^p_\ell \otimes S^q_m
\end{equation}

with $0\le p<\ell$, $0\le q<m$. A bivariate bicycle code is defined by a pair of matrices 
\begin{equation}
    A = A_1 + A_2 + A_3 \ \mbox{and} \ B = B_1 + B_2 + B_3
\end{equation}

where each term $A_i$ and $B_j$ is a power of $x$ or $y$ \cite{Bravyi2024}. The check matrices for the code are given by
\begin{equation}
    H^X = \left[A | B\right] \ \mbox{and} \  H^Z=\left[B^T | A^T\right].
\end{equation} 
As an explicit example, with $\ell=12$ and $m=6$, the matrices $A=x^3+y+y^2$ and $B=y^3+x+x^2$ result in the [[144,12,12]] code from Ref.~\cite{Bravyi2024}. In this way, an $X$ check represented by row $i$ involves a data qubit represented by column $j$ if $H^X_{ij}=1$, and likewise for $Z$ checks and $H^Z$. We follow the labeling convention from Ref.~\cite{Bravyi2024} by partitioning the data qubits into two sets $L$ and $R$ given by the left and right blocks of columns in the check matrices, respectively. This gives four sets of qubits $L$, $R$, $X$, and $Z$, each of size $n/2=\ell m$, which we may assign integer labels from $\mathbb{Z}_{\ell m}=\{0,1,\dots,\ell m-1\}$. Data qubits and checks may equivalently be labeled by monomials from $\mathcal{M} =\{1,y,\dots,y^{m-1},x,xy,\dots,xy^{m-1},\dots,x^{\ell-1}y^{m-1}\}$ so that a qubit with label $i\in\mathbb{Z}_{\ell m}$ may alternatively be labeled by $x^{a_i}y^{i-ma_i}$ for $a_i=\text{floor}(i/m)$ \cite{Bravyi2024}. Our goal is to assign each qubit a position on a grid with open boundary conditions such that the distance $D_{\rm max}$ of the longest range check operation is minimized.

A brute force search of all possible layouts is not feasible due to the combinatorially large number of possibilities. Instead, we perform a limited search over initial position assignments for the $L$ and $R$ data qubits, and for each initial layout we solve for the assignment of the ancilla $X$ and $Z$ qubit positions which minimizes the distance $D_{\rm max}$ of the longest range check operation.  Note that the set $\mathcal{M}$ together with poloynomial multiplication forms an abelian group isomorphic to $\mathbb{Z}_l \times \mathbb{Z}_m$. We seek a one to one mapping of elements from this group onto a rectangular grid. The mapping we use is defined by five monomials $L_1$, $L_2$, $R_1$, $R_2$, $LR \in \mathcal{M}$. For the mapping to be well defined, $L_1$ and $L_2$ must satisfy conditions of lemma 4 from Ref.~\cite{Bravyi2024} to obtain a well defined enumeration of the $L$ qubits:
\begin{enumerate}[label=(\roman*)]
\item $\langle L_1, L_2\rangle=\mathcal{M}$ and
\item $\ord{L_1}\ord{ L_2}=\ell m$.
\end{enumerate}
Here, $\langle L_1, L_2 \rangle$ is the group generated by the subset $\{L_1, L_2\}$ and $\ord{\alpha}$ is the order of $\alpha$, the smallest positive integer $p$ such that $\alpha^p=1$. $R_1$ and $R_2$ must also satisfy these conditions, with the additional constraint that $\ord{R_1}=\ord{L_1}$ and $\ord{R_2}=\ord{L_2}.$ Letting $\mu=\ord{L_1}$ and $\lambda=\ord{L_2}$, we construct a $2\mu \times 2\lambda$ grid of qubits. For each $L$ qubit with label $\alpha_L \in \mathcal{M}$, there is a unique $(a, b) \in \mathbb{Z}_\mu \times \mathbb{Z}_\lambda$ such that $\alpha_L = L_1^aL_2^b$. We initially assign qubit $L$ with label $\alpha_L$ to the grid position $(2a, 2b)$. Likewise, there is a unique $(a, b) \in \mathbb{Z}_\mu \times \mathbb{Z}_\lambda$ such that $\alpha_R = LR(R_1^aR_2^b$) and initially assign data qubit $R$ with label $\alpha_R$ to grid position $(2a+1,2b+1)$. The layout of data qubits from \cite{Bravyi2024} and \cite{Viszlai2023} may be reproduced as special cases by the appropriate choice of five monomials. In \cite{Bravyi2024}, monomials $A_i$, $A_j^T$, $B_g$, and $B_h^T$ are selected from the monomial terms that produce the code to form $L_1=R_1=A_iA_j^T$, $L_2=R_2=B_gB^T_h$, $LR=A_j^T B_g$. By selecting the monomials in this way, they produce a ``toric" layout where, if the grid is considered to have periodic boundary conditions, four of the six connections for each qubit are nearest neighbor interactions. To produce the layout of \cite{Viszlai2023}, take $L_1=R_1=y$, $L_2=R_2=x$, $LR=1$, which directly enumerates the qubits in a column-major order. 

We search over all possible values of $L_1$ and $L_2$ which satisfy the above conditions. For the smaller codes $(n < 144)$, we also searched over all values of $R_1$ and $R_2$, but found that the best result consistently had $R_{1(2)} \in \{ {L_{1(2)}, L^T_{1(2)}} \}$. Thus, we limit the search for larger codes to this range. Lastly, we let $LR \in \mathcal{M}$ be arbitrary. After initially placing each of the $L$ and $R$ qubits, the entire grid is remapped according to the plane ``folding" procedure illustrated in Fig. 2 in the main text.  At this stage, the positions of $L$ and $R$ qubits are considered fixed, and we must identify an assignment of $X$ and $Z$ qubits which minimizes the maximum check operator distance $D_{\rm max}$.
This is done using a bipartite matching procedure as explained in the main text. 

\begin{figure*}[!t]
\centering
 \includegraphics[width=.95\textwidth]{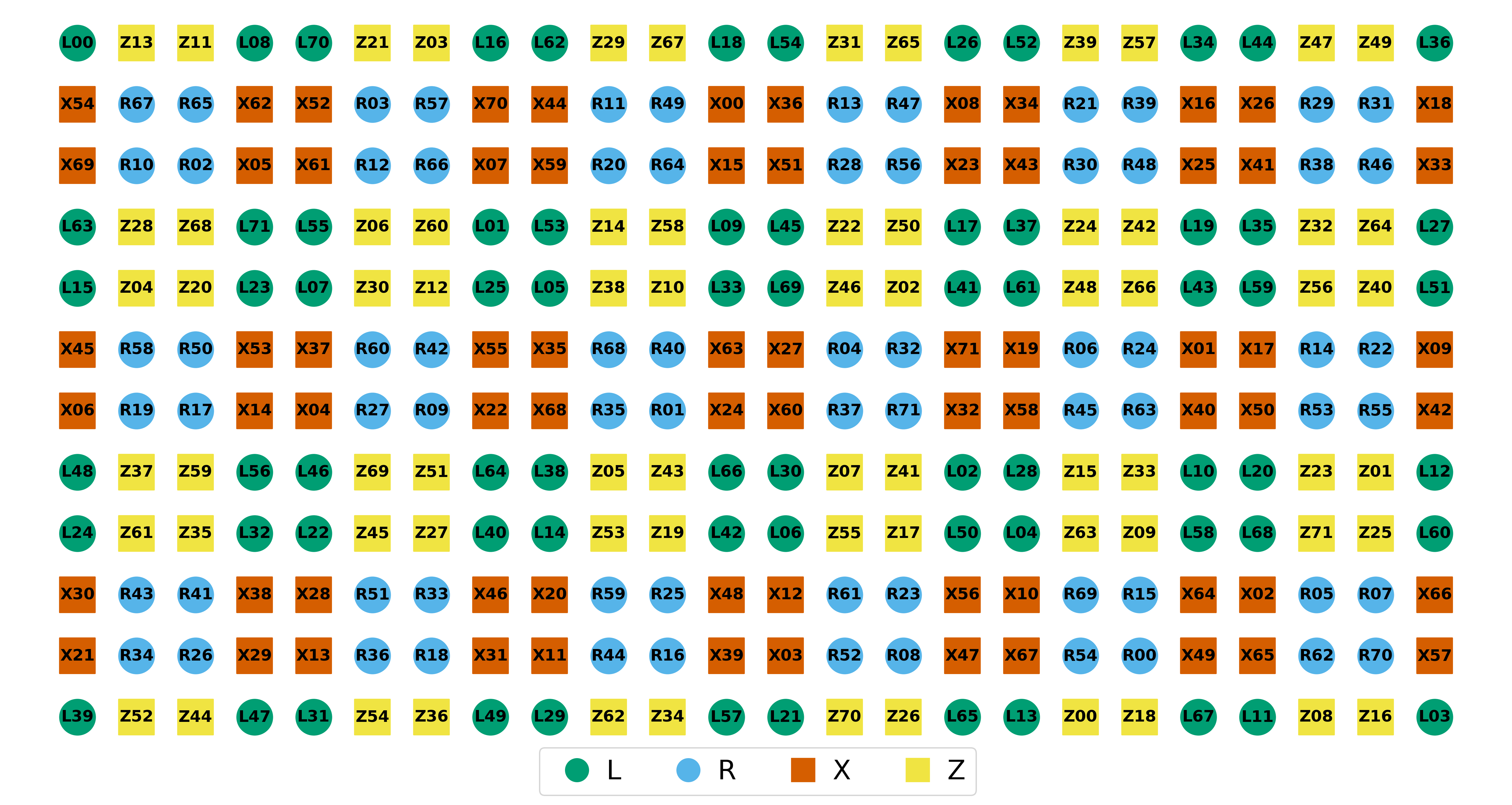}
 \vspace{-.5cm}
  \caption{Optimized layout of the [[144,12,12]] code with a maximum communication distance of 7.21 lattice spacings.  All nodes are labeled. Edges can be generated via the rules for the [[144,12,12]] code.}
\label{fig.sup_layout}
\end{figure*}

As an example the  layout of the [[144,12,12]] code after optimization to achieve $D_{\rm max}=7.21$ is given in Fig. \ref{fig.sup_layout}. A table enumerating the monomials that produce the best layouts found for each code is given in Table \ref{tab.qLDPCWithMonomials}.

\begin{table*}
\caption{
Bivariate bicycle codes with $n$ data qubits, $2n$ total qubits (including ancillae), $k$ logical qubits, and code distance $d$. For each code, we provide the parameters defining the code \cite{Bravyi2024}, the monomials used to generate the optimal layout, and the maximum communication distance $D_{\rm max}$ in units of the lattice spacing using our method.
}
\label{tab.qLDPCWithMonomials}
\centering
\begin{tabular}{|c| c| c|c|c|c|c|c|c|c|c}
\hline
$[[n,k,d]]$     & $\ell,m$  &A           & B      & $L_1$ & $L_2$ & $R_1$ & $R_2$ & $LR$&$D_{\rm max}$ \\
\hline
$[[72,12,6]]$   & 6, 6  &$x^3+y+y^2$   & $y^3+x+x^2$  & $x^2y$  & $x^5y^5$  & $x^4y^5$  &  $xy$ & $x^3$ & 5. \\
$[[90,8,10]]$   & 15, 3 &$x^9+y+y^2$   & $1+x^2+x^7$  & $x^5$ & $x^9y$  &  $x^{10}$ & $x^6y^2$ &  $x$ & 10.\\
$[[108,8,10]]$  & 9, 6  &$x^3+y+y^2$   & $y^3+x+x^2$  & $x^3y$  & $x^7y^2$ & $x^3xy$&   $x^2y^4$  & $x^5$ & 7.    \\
$[[144,12,12]]$ & 12, 6 &$x^3+y+y^2$   & $y^3+x+x^2$  & $x^2y^3$ & $x^{11}y^4$ & $x^2y^3$  & $x^{11}y^4$  & $y^2$ & 7.21  \\
$[[288,12,18]]$ & 12, 12&$x^3+y^2+y^7$ & $y^3+x+x^2$  & $y^7$  & $xy^9$  & $y^7$ &  $xy^9$ & $y^6$ & 7.21  \\
\hline
\end{tabular}
\end{table*}

\section{Concurrency of gate operations}
\label{sec.App.concurrency}

Due to the long range nature of the Rydberg interaction only a limited number of gates can be executed in one clock cycle. \rsub{We performed simulations of the sensitivity of the {\sf CZ} gates to blockade crosstalk as a function of $x_i = \Delta_{{\rm B}i} t_{\rm gate} / (2 \pi)$, where $\Delta_{{\rm B}i}$ is the sum of the interaction strengths between atoms not in the same {\sf CZ} pair as qubit $i$ with qubit $i$. For all gate distances, if both qubits satisfy $x_i < 0.01$, then the infidelity due to blockade leakage is upper bounded by $10^{-4}.$ The full check operation circuit is divided into two subcircuits, the first of which performs the {\sf CZ} gates for the {\sf X} checks and the latter the {\sf CZ} gates for the {\sf Z} checks. These subcircuits are sectioned off from each other by single qubit rotations which prevent {\sf CZ} operations from each subcircuit from operating simultaneously.} For the layout in Fig. \ref{fig.sup_layout}, every check  in rows \rsub{12}
-17 of Table \ref{tab.distances2} interferes with all other checks of the same distance \rsub{in the same subcircuit}, and thus they must be performed one at a time. \rsub{{\sf CZ} operations in the remaining rows can be at least partially parallelized according to this constraint}. 

For each of these rows \rsub{and in each subcircuit}, we apply a greedy algorithm which selects a grouping of maximum size to be scheduled. Any other groupings of maximum size that involved qubits in the selected group are removed from consideration, and the process is repeated until every check operation is included in one of the groupings. During this process, the maximum cardinality of the remaining groupings may decrease. The specific choice of maximum size grouping to be selected is determined by choosing the option which maximizes the number of remaining groupings. If there are multiple equivalently good options to choose from based on this metric, we may choose at random and repeat this process several times, keeping the best end result. 

By applying this process for each row in Table \ref{tab.distances}, we get a total of 
\rsub{693} clock cycles \rsub{dedicated to the {\sf CZ} gates, with a total {\sf CZ} gate time of ~$234~\rm\mu s$. Single qubit gates are necessary to perform the basis transformations to convert the {\sf CZ} gates to {\sf CNOT} gates as well as for qubit measurement. A global microwave field is used when all qubits can be rotated with the same operation, and parallel Raman gates are used when only certain subsets of qubits require an operation. Parallelized local {\sf Z} gates correct any local phase shift induced on the atoms during the {\sf CZ} gates. In total, these local gates add a fixed cost gate time of approximately $3.5~\rm \mu s$, and the parallelized local phase correction gates add four cycles worth of switching time. This leads to a fixed time cost of $237.5~\rm\mu s$ and a variable time cost equal to $693+4=697$ times the switching time. Assuming a switching time of $1.5~\rm\mu s,$} this leads to the estimate for a full QEC cycle \rsub{of 1.28~\rm ms} given in the main text. 

We note that the number of clock cycles could be  reduced somewhat by introducing additional Rydberg levels such that the range of the Rydberg interaction is as short as possible consistent with performing a gate of high fidelity at the desired distance.

\section{Gate fidelity}
\label{sec.App.fidelity}

Consider a generalized diagonal gate  $\sf U_{\rm d}$  that results in the  unitary evolution   
\bea
\ket{00}&\rightarrow&{\sf U}_{\rm d}\ket{00} =a_{00} e^{\imath \phi_{00}}\ket{00}\nonumber\\
\ket{01}& \rightarrow&{\sf U}_{\rm d}\ket{01} =a_{01}e^{\imath\phi_{01}}\ket{01}\nonumber\\
\ket{10}&\rightarrow& {\sf U}_{\rm d}\ket{10} =a_{10}e^{\imath\phi_{10}}\ket{10}\nonumber\\
\ket{11}&\rightarrow&{\sf U_{\rm d}}\ket{11} = a_{11}e^{\imath\phi_{11}}\ket{11}
\label{eq.Udiagonal}
\eea
where $a_{ij}$ are real amplitudes and $\phi_{ij}$ are real phases. A phase gate would have $a_{00}=a_{01}=a_{10}=a_{11}=1.$
Imperfect Rydberg phase gates have the $a_{ij}\le 1$ due to rotation errors that do not fully return population to the computational basis states. 

A   Bell state can be  prepared with the sequence  
\begin{widetext}
\be
\ket{\psi}=[\sfI\otimes \sfR_{x}(\pi/2)][\sfR_z(-\phi_{10})\otimes \sfR_z(-\phi_{01})]\sfU_{\rm d}[\sfR_x(\pi/2)\otimes \sfR_x(\pi/2)]\ket{00}.
\label{eq.psiBell}
\ee
\end{widetext}
In the ideal case where $a_{00}=a_{01}=a_{10}=a_{11}=1$, $\phi_{00}=0,$ and introducing the gate phase 
$\phi=\phi_{11}-\phi_{01}-\phi_{10}$ the output state is
$$
\ket{B}=-i e^{\imath \frac{\phi_{01}+\phi_{10}}{2}}\frac{\ket{01}+\ket{10}}{\sqrt2}
$$
when $\phi=\pi$.
This is a maximally entangled Bell state up to an irrelevant global phase. 

In the phase gate limit with $a_{ij}=1$ the Bell  fidelity of the output state is 
\bea
{\mathcal F}_{\rm pg}&=&|\bra{B}\psi\rangle|^2\nonumber\\
&=&\frac{3+2\cos(\phi_{00})-\cos(\phi-\phi_{00})-2\cos(\phi)}{8}.
\label{eq.FBell}\eea
The Bell fidelity depends on the phase $\phi_{00}$.
For Rydberg gates that couple state $\ket{1}$ to the Rydberg level off-resonant light shifts are responsible for the phase  $\phi_{00}$.  These shifts can be cancelled by applying a sequence of $z$ and $x$ single qubit rotations. 
We may therefore  set $\phi_{00}=0$ in which case the fidelity  reduces to
$$
{\mathcal F}_{\rm pg,0}=\frac{5-3\cos(\phi)}{8}.
$$

For realistic Rydberg gates there is also an error contribution due to imperfect rotations that leave population outside the qubit basis. To analyze the fidelity in this situation we  work in a three dimensional basis $\{\ket{0},\ket{1},\ket{r}\}$, with $\ket{r}$ the Rydberg state.  Tracing over the Rydberg state,   we have an effective unitary in the qubit basis of the form of  eq. (\ref{eq.Udiagonal}). 

Assuming  $a_{00}=1$ and a symmetric gate such that $a_{10}=a_{01}, \phi_{10}=\phi_{01}$ we find  
\begin{widetext}
\be
{\mathcal F}_{\rm sym}=\frac{1+4 a_{01}^2 + a_{11}^2 +4 a_{01}\cos(\phi_{00})-2 a_{11}\cos(\phi-\phi_{00})-4 a_{01} a_{11}\cos(\phi)}{16}.
\ee
\end{widetext}
If we again assume $\phi_{00}=0$ this reduces to
 \be
{\mathcal F}_{\rm sym,0}=\frac{1+4 a_{01}^2 +4 a_{01}+ a_{11}^2 -2(1+2a_{01}) a_{11}\cos(\phi)}{16}.
\label{eq.Fsym0}
\ee

Equation (\ref{eq.Fsym0}) gives the fidelity for preparing a Bell state starting from the  state $\ket{00}$. We can generalize this to give the fidelity of an operator $\sfU$ with respect to the ideal operator $\sfU_0$ averaged over all two-qubit initial states.  A convenient expression for calculating the average fidelity is \cite{Pedersen2007}
\bea 
{\mathcal F}_{\rm ave}&=&\frac{n+\left|\textrm{Tr}(\sfU_0^\dag \sfU) \right|^2 }{n(n+1)}.\nonumber
\eea
where $n$ is the dimension of the Hilbert space and we have used the fact that $\sfU_0$ and $\sfU$ are unitary. Using $n=4,$ 
\bea
\sfU_0&=&\begin{pmatrix}
1& 0 & 0 &0 \\ 0 & 1 & 0 & 0 \\ 0 & 0 & 1 & 0 \\ 0 & 0 & 0 & -1
\end{pmatrix},\nonumber\\
\sfU&=&[\sfR_z(-\phi_{10})\otimes \sfR_z(-\phi_{01})]\sfU_{\rm d}\nonumber\\
&=&e^{\imath \frac{\phi_{01}+\phi_{10}}{2}}\begin{pmatrix}
a_{00}e^{\imath \phi_{00}} & 0 & 0 &0 \\ 0 & a_{01}  & 0 & 0 \\ 0 & 0 & a_{10} & 0 \\ 0 & 0 & 0 & a_{11}e^{i \phi}
\end{pmatrix}
\eea and assuming $a_{00}=1, \phi_{00}=0$ with the symmetric conditions $a_{10}=a_{01}, \phi_{10}=\phi_{01}$, we find  
\bea 
{\mathcal F}_{\rm ave}&=&\frac{5+4 a_{01}^2 + 4 a_{01} + a_{11}^2 -2 (1+2 a_{01} )a_{11}\cos(\phi)}{20}.\nonumber\\
\label{eq.CZaveragefidelity}
\eea   

The average fidelity ${\mathcal F}_{\rm ave}$ was found by simulating the coherent evolution using the gate and interaction parameters from Table \ref{tab.distances2} by numerical solution of  the Schr\"odinger equation and extracting the values of $a_{01}, a_{11},$ and $\phi$. The fidelity reported in the main text and in Table \ref{tab.distances2} was corrected for Rydberg state decay using  
\be
{\mathcal F}={\mathcal F}_{\rm ave}- T_{\rm R}/\tau_{\rm R}
\ee
where $\tau_{\rm R}$ is the Rydberg lifetime and the integrated Rydberg population averaged over the computational basis states is 
$$
T_{\rm R}=\frac{1}{2}\int_0^{t_{\rm gate}} dt\, 
|c_{\rm R}(t)|^2+|c_{\rm gR}(t)|^2+|c_{\rm RR}(t)|^2.
$$
Here $c_{\rm R}(t)$  is the Rydberg amplitude when one atom is Rydberg coupled, 
$c_{\rm gR}(t)$  is the amplitude of the singly occupied Rydberg state when two atoms are Rydberg coupled, and  $c_{\rm RR}(t)$ is the amplitude of the doubly occupied Rydberg state.  All calculations assumed symmetrical behavior of the two atoms. 

\section{Analytical gate parameters}
\label{sec.App.gate}

The analytical pulses are based on a constant amplitude Rabi drive with smooth turn on and turn off and a modulated phase profile. The Rabi amplitude is given by  
\bea
\Omega(t)&=&\Omega_0\left[\frac{1}{1+e^{-(t-20\tau_{\rm e})/\tau_{\rm e}}}\right. \nonumber \\
&&\left.~~~ +\frac{1}{1+e^{-(t_{\rm gate}-20\tau_{\rm e} -t)/\tau_{\rm e}}} -1\right]
\eea
with $\tau_{\rm e}=1.825~\rm ns$ and $t_{\rm gate}$ the nominal gate duration. The phase profile is 
\be
\phi(t)=\Delta_0 t + a\sin[2\pi f(t-t_0)]e^{-[(t-t_0)/\tau]^4}
\label{eq.phase}
\ee
with $t_0=t_{\rm gate}/2$ the midpoint. The parameters  
$\Omega_0, \Delta_0, a, f, \tau$ are chosen to optimize the fidelity for given $V, \tau_R$ values.  
Table \ref{tab.distances2} provides parameters found from numerical search for the analytical Rydberg gate designs used in the main text.

 \begin{table*}[!t]
\caption{
Gate parameters and performance for the  Euclidean communication distances $D$ in units of lattice spacing for the 864 check operations in the   $[[144,12,12]]$ code. The interaction strength is given at  $R=1.7 D~ (\mu\rm m)$. The lifetimes of the Rydberg levels are $\tau_{\rm R}=60.4, 209, 252~(\mu\rm s)$ for $n=50,83,90$ and the indicated fidelity includes only rotation and Rydberg lifetime errors.} \label{tab.distances2}
\centering
\begin{tabular}{|c| c| c|c|c|c|c|c|c|c|c|c|c|c|}
\hline
gate&lattice  & connection  & Rydberg & interaction & fidelity & $t_{\rm gate}$ & $a$ & $f$ & $\Omega/2\pi$ & $\Delta_0/2\pi$ & $\tau$& $V/\Omega$ & $\epsilon/\epsilon_{\rm min}$ \\
&distance $D$& $R~(\mu\rm m)$& level& 
$V(R)/2\pi$ (MHz)  & $\mathcal F$& (ns) &  & (MHz) & (MHz) & (MHz) & (ns)& & \\
\hline
1&1 & 1.7 & $50s_{1/2}$&415. & 0.9996& 130 & 0.774 & 20.0 & 21.5 & -1.59 & 1907&19.3&31.\\
2&1.41 &2.40 & $50s_{1/2}$&58.5& 0.9993& 180 & 0.749 & 10.6 & 11.3& -1.59 & 374&5.2&7.8\\
\hline
3&2.0 & 3.4 & $83s_{1/2}$&1160.& 0.9999& 150 & 1.44 & 9.96 & 15.9 & -0.818& 20.3&73.&76.\\
4&2.24 & 3.81&$83s_{1/2}$&780.& 0.9999& 150 & 1.45 & 9.91 & 15.9 & -0.856& 20.2&49.&51.\\
5&3.16 &5.37 & $83s_{1/2}$&170. & 0.9998 & 180 & 0.707 & 11.5 & 11.4 & -0.451& 744&15. &22.\\
6&3.61&6.14&$83s_{1/2}$&85.& 0.9998 & 180 & 0.594 & 13.3 & 11.2 & 0.152 & 922&7.6&11.\\
7&4.12 &7.00 & $83s_{1/2}$&40. & 0.9998 & 180 & 0.569 & 14.1 & 11.1 & -0.505& 81&3.6&5.3\\
8&4.47&7.60&$83s_{1/2}$&25.& 0.9998& 200 & 0.622& 14.4& 9.54& -0.897 & 452& 2.6& 3.3\\
9&5 & 8.65& $83s_{1/2}$&13. & 0.9996& 270 & 0.439 & 14.46 & 11.2 & 0.395& 97.3&1.2&3.4\\
10&5.10 & 8.67& $83s_{1/2}$&11.5 & 0.9996& 270 & 0.502 & 14.63& 11.545 & 0.34& 94.2&1.0&3.0\\
11&5.83 & 9.91& $83s_{1/2}$&5.2 & 0.9995& 270 & 2 & 12.9& 6.72 & -0.976& 34.5&0.77&1.7\\
\hline
12&6 & 10.2& $90s_{1/2}$&11.2 & 0.9996& 350 & 0.578 & 6.42 & 4.32 & -0.383&1756&2.6&3.5\\
13&6.08 & 10.3& $90s_{1/2}$&10.1 & 0.9995& 400 & 0.725 & 5.89 & 3.85 & -0.469& 821 &2.6&4.0\\
14&6.40 & 10.9& $90s_{1/2}$&7.7 & 0.9994& 450 & 0.496 & 8.01 & 5.88 & 0.422& 197&1.3&3.7\\
15&6.71 & 11.4& $90s_{1/2}$&5.8 & 0.9994& 460 & 0.556 & 7.54 & 5.98 & 0.338& 171&0.97&2.8\\
16&7.07&12.0&$90s_{1/2}$&4.3&0.9993& 465 & 1.16 & 5.05 & 6.36 & 0.794 & 367&0.68&2.4\\
17&7.21 & 12.3& $90s_{1/2}$&3.8& 0.9993& 480 & 1.46 & 7.06 & 7.26 & 0.074& 103&0.52&2.1\\
\hline
\end{tabular}
\end{table*}

As can be seen in the Table \rsub{\ref{tab.distances2}} it is possible to choose parameters that provide fidelity ${\mathcal F}\ge 0.9993$ for inter-atomic distances ranging from 1.7 to 12.3 $\mu\rm m$. The gate designs have a wide range of parameters with the apodization parameter $\tau$ in eq. (\ref{eq.phase}) being much less than $t_{\rm gate}$ in some case and much larger than $t_{\rm gate}$ in others. When $\tau\gg t_{\rm gate}$ the phase function reduces to that used in \cite{Evered2023}. None of the gates are in the strong blockade regime with $V/\Omega$ ranging from 73. all the way down to 0.52. This implies that each gate must be calibrated for specific inter-atomic distances.

The reported fidelities are for the idealized situation of no additional technical errors due to laser noise, optical beam pointing,
atom motion and position variations at
finite temperature, or 
electric and magnetic field noise. Pulse designs that suppress sensitivity to these effects \cite{Saffman2020,Mohan2023,Jandura2023,Fromonteil2023,WXLi2024} are important for implementation, but are outside the scope of this work. \rsub{Additional error sources arise due to dipole-dipole forces during double excitation to the Rydberg state and Doppler broadening of the Rydberg transition \cite{Robicheaux2021}. While these error sources can be largely mitigated by cooling the atoms to the ground state of the traps, they still have a significant effect on fidelity when the blockade is small. We have calculated gate infidelities due to dipole-dipole repulsion and Doppler broadening in Table \ref{tab.dipole_dipole}. Including these additional error terms all 17 sets of $\sf CZ$ gate parameters still have $\mathcal F\ge 0.999$. }\\

 \begin{table*}[!t]
\caption{
\rsub{Simulated gat fidelities including the contributions from dipole-dipole forces due to  double Rydberg excitation ($\epsilon_{\rm RR}$) and the Doppler error ($\epsilon_{\rm Doppler}$) for the gate parameters listed in table \ref{tab.distances2}. The error due to double excitation has been calculated using the methods described in \cite{Robicheaux2021} and depends on the distance that the two atoms are separated ($R$), the interaction strength ($V(R)$), and the effective time the atoms spend in the doubly excited Rydberg state given an initial state of $\ket{11}$ ($T_{\rm RR}$).  Doppler broadening of the Rydberg transition  leads to a gate error ($\epsilon_{\rm Doppler}$) which  depends on the integrated time the atoms spend in the Rydberg state, $T_{\rm R}$, and the momentum distribution of the atoms \cite{Robicheaux2021}.  These errors lead to modified fidelity ($\mathcal F'$) and infidelity ($\epsilon'$) metrics. We have assumed ground state cooling into a trap with a radial frequency of $\omega_r=2\pi\times 100$ kHz. It is possible to further reduce $\epsilon_{RR}$ by using higher radial trap frequency; however, reducing this error source to a negligible level requires a prohibitively high trap frequency when $V/\Omega$ is not much larger than unity. For example, a trap frequency of $\omega_r=2\pi\times 1.25$ MHz would be needed for gate 17 to maintain $\epsilon'/\epsilon_{\rm min}=2.1$ }} \label{tab.dipole_dipole}
\centering
\begin{tabular}{|c| c|c| c|c|c|c|c| c|}
\hline
gate& connection $R~(\mu\rm m)$ & $V/\Omega$& $T_{\rm RR}$ (ns) & $\epsilon_{RR}$ & $T_{\rm R}$ (ns) & $\epsilon_{\rm Doppler}$ & fidelity $\mathcal F'$  & $\epsilon' / \epsilon_{\rm min}$\\
\hline
1& 1.7 & 19.3 & 0.04 &  $1.5 \times 10^{-5}$ & 21 & $8.2 \times 10^{-7}$&0.9996 & 31.\\
2 & 2.4& 5.2 & 1.3 &  $1.7 \times 10^{-4}$ & 39 & $2.70 \times 10^{-6}$ & 0.9991 & 10\\
\hline
3 & 3.4& 73 & 0.004 &  $2.7 \times 10^{-7}$ & 30 & $1.6 \times 10^{-6}$& 0.9999 & 76.\\
4 & 3.81& 49 & 0.008 &  $4.9 \times 10^{-7}$ & 30 &$1.6 \times 10^{-6}$ & 0.9999 & 51.\\
5 & 5.37& 15 & 0.13 &  $2.7 \times 10^{-6}$ & 40&$2.9 \times 10^{-6}$ & 0.9998 & 22.\\
6 & 6.14 & 7.6 &  0.53 &  $8.8 \times 10^{-6}$ & 40 & $2.9 \times 10^{-6}$ & 0.9998 & 11\\
7 & 7.00 & 3.6&  3.4 &  $6.4 \times 10^{-5} $ &40 &$2.8 \times 10^{-6}$ & 0.9997 & 8.0\\
8 & 7.60 & 2.6& 14 &  $3.7 \times 10^{-4}$ & 48 & $4.0 \times 10^{-6}$& 0.9994 & 10\\
9 & 8.65 & 1.2& 24 &  $2.1 \times 10^{-4}$ & 81& $1.2 \times 10^{-5}$& 0.9994 & 5.0\\
10 & 8.67 & 1.0& 29 &  $2.5 \times 10^{-4}$ &83 & $1.2 \times 10^{-5}$& 0.9993 & 5.3\\
11 & 9.91 & 0.77& 89 &  $3.6 \times 10^{-4}$ &111 &$2.2 \times 10^{-5}$ & 0.9991 & 3.1\\
\hline
12 & 10.2 & 2.6& 32 &  $1.9 \times 10^{-4}$ &104 &$1.9 \times 10^{-5}$ & 0.9994 & 5.3 \\
13 & 10.3 & 2.6& 40 &  $2.6 \times 10^{-4}$ & 122&$2.6 \times 10^{-5}$ & 0.9992 & 6.4\\
14 & 10.9 & 1.3& 53 &  $2.2 \times 10^{-4}$ &157 & $4.3 \times 10^{-5}$ & 0.9991 & 5.6\\
15 & 11.4 & 0.97& 60 &  $1.5 \times 10^{-4}$ & 162& $4.6 \times 10^{-5}$& 0.9992 & 3.7\\
16 & 12.0 & 0.68& 82 &  $1.4 \times 10^{-4}$ & 184 & $5.9 \times 10^{-5}$ & 0.9991 & 3.1\\
17 & 12.3 & 0.52& 120 &  $2.2 \times 10^{-4}$ & 183 & $5.9 \times 10^{-5}$ & 0.9990 & 3.0\\
\hline
\end{tabular}
\end{table*}

\end{document}